# Highly sticky surfaces made by electrospun polymer nanofibers

S. Varagnolo[a], F. Raccanello[a], M. Pierno[a], G. Mistura*[,a],
M. Moffa[b], L. Persano[b], D. Pisignano*[,b,c]

[a] CNISM and Dipartimento di Fisica e Astronomia "G. Galilei", Università di Padova, via Marzolo 8, I-35131 Padova, Italy

[b] NEST, Istituto Nanoscienze-CNR, Piazza S. Silvestro 12, I-56127 Pisa, Italy

[c] Dipartimento di Matematica e Fisica "Ennio De Giorgi", Università del Salento, via Arnesano, I-73100, Lecce, Italy

Email: G. Mistura: email, giampaolo.mistura@unipd.it

D. Pisignano: email, dario.pisgnano@unisalento.it

## Abstract

We report on a comprehensive study of the unique adhesive properties of mats of polymethylmethacrylate (PMMA) nanofibers produced by electrospinning. Fibers are deposited on glass, varying the diameter and the relative orientation of the polymer filaments (random *vs* aligned configuration). While no significant variation is observed in the static contact angle (~130°) of deposited water drops upon changing the average fiber diameter up to the micrometer scale, fibers are found to exhibit unequalled water adhesion. Placed vertically, they can hold up water drops as large as 60 μL, more than twice the values typically obtained with hairy surfaces prepared by different methods. For aligned fibers with anisotropic wetting behavior, the maximum volume measured in the direction perpendicular to the fibers goes up to 90 μL. This work suggests new routes to tailor the wetting behavior on extended areas by nanofiber coatings, with possible applications in adsorbing and catalytic surfaces, microfluidic devices, and filtration technologies.

## 1. Introduction

Wettability is one of the most important properties of solids, affecting their surface mechanics, tribology, resistance, and biocompatibility, and being governed by both the chemical composition and the morphology of the involved interface.[1, 2] In this respect, surfaces that have attracted a lot of attention in recent years are those exhibiting superhydrophobicity, namely an apparent static contact angle, $\theta$, formed by water drops greater than 150° inspired by many plants and insects, which provides new and versatile ideas for designing materials with self-cleaning and antifouling properties, and drag reduction.[3] However, the dynamic behavior of these bioinspired surfaces can vary significantly.[4-6] On a lotus leaf, water drops roll off very easily even at inclination well below 10°, removing dust particles present on the surface (self-cleaning or lotus effect).[7] In contrast, large water drops stick to rose petals even though they are tilted upside down (petal effect).[8] Key features to achieve a specific superhydrophobic behavior involve both a proper chemical composition of the surface and an appropriate roughness at the micro/nanometer scale,[2, 9, 10] since in untextured surfaces $\theta$ is generally below ~120°, which is the value characteristic of fluorinated materials.[1]

Various physical and chemical methods have been employed to realize either self-cleaning[11-15] or sticky[8, 16-19] superhydrophobic surfaces. In this framework, electrospinning provides a simple and practical way to tailor surface roughness over large areas through coatings made of fibers with diameters ranging from tens of µm to tens of nm, which are produced from polymer solutions with sufficient molecular entanglements.[20-22] To this aim, a high voltage is applied to the solution, which is extruded from a spinneret as an electrified jet.[23] The resulting materials, in a variety of forms ranging from individual nanofibers to non-woven mats with large area, have found application in many fields, including the realization of self-cleaning, superhydrophobic and superolephobic coatings.[24-35] Their wettability has been mainly assessed by measuring the apparent contact angle ($\cong 150°$) and the roll off angle ($\leq 10°$), and explained in terms of the standard Cassie model with the drop contacting a composite landscape of trapped air and solid substrate.[1]

Here, we focus on a different property, studying highly sticky hydrophobic fibers, and investigate a feature scarcely addressed[36-38] hitherto. The large adhesion is attributed to the water drop partially penetrating the surface texture according to a Cassie impregnating model.[1] We electrospin polymethylmethacrylate (PMMA) fibers onto different substrates, varying the diameter and the relative orientation of the polymer filaments (random *vs* aligned configuration). Very large adhesion forces, capable to hold water drops as large as 90 μL are found. To better understand the role played by the substrate, the wetting data are compared with those obtained on free-standing mats. This study suggests new routes to produce coatings with tailored wetting properties which can easily cover extended surface areas. Potential applications of these findings include the design and the fabrication of new adsorbing media, catalytic surfaces, delivery of fluids with reduced or no volumetric loss, lab-on-chip architectures, and materials supporting water remediation.

## 2. Methodology

**Electrospinning**

Fibers are electrospun from a chloroform solution of PMMA (120,000 g/mol, Sigma-Aldrich). The electrospinning apparatus consists of a 1 mL syringe tipped with a 27-gauge steel needle, mounted on a pump providing a flow rate of 0.5 mL/h (33 Dual Syringe Pump, Harvard Apparatus Inc.). The syringe needle is biased at 10 kV by a high voltage power supplier (EL60R0.6-22, Glassman High Voltage). Randomly oriented fibers are deposited on (18×18 mm$^2$) glass slides, positioned at 15 cm from the needle, whereas uniaxially aligned fibers are realized by positioning the substrates on a disk (8 cm diameter, 1 cm thickness) rotating at 4000 rpm. Free-standing non-wovens made of nanofibers, with overall thickness of about 60 μm, are also produced by depositing fibers over a time up to three hours.

Fibers are inspected by atomic force microscopy (AFM, multimode head equipped with a Nanoscope IIIa electronic controller, Veeco), and by scanning electron microscopy (SEM, Nova NanoSEM 450, FEI) following the thermal deposition of 5 nm of Cr. AFM micrographs are acquired in tapping mode, using Sb-doped Si cantilevers with resonance frequency of 76 kHz. The average diameter ($\phi$) of the fibers, calculated from at least 100 filaments per each species imaged by SEM, is found to range from 0.6 to about 5.2 µm, with overall substrate coating thickness of about 10 µm. Fibers with diameters well below the micrometer-scale (down to $\phi$=0.6 µm) are spun by adding the organic salt tetrabutylammonium iodide (TBAI, Sigma-Aldrich) to the solution, at a concentration of 1% wt/wt relative to PMMA. In the following, samples labeled as *R* and *A* refer to randomly distributed and to aligned fibers, respectively, whereas progressive numbers from *1* to *4* indicate increasing average diameters (*1*: $\phi$=0.6-0.7 µm, *2*: $\phi$=1.2-1.3 µm, *3*: $\phi$=1.7-1.8 µm, and *4*: $\phi$=4.6-5.2 µm, respectively), as reported in Tables 1 and 2.

**Wettability characterization**

Measurements of the apparent contact angle ($\theta$) are performed depositing 1 µL drops of distilled water on samples positioned horizontally. Data are collected from at least five drops placed in different positions, and the final contact angle is expressed as the mean of the various values, with error given by the standard deviation which is representative of the surface uniformity. For aligned fibers, drops are simultaneously viewed by two high-resolution cameras from orthogonal directions: parallel and perpendicular to the fibers orientation. The cameras mount telecentric lenses that guarantee good contrast and faithful imaging of the drop which is illuminated by two back-light collimated LED sources.[39] The profile of each image is analyzed off-line by using a custom made program.[40] For each drop, the apparent contact angle is deduced from a fit of the profile.

Advancing ($\theta_A$) and receding ($\theta_R$) contact angles are measured through the dynamic sessile drop method by injecting and removing water from a drop having an initial volume of almost 1 µL and are defined as the threshold angles necessary to observe an expansion and a contraction of the contact line, respectively. For each surface, experiments are repeated in at least three different positions. From these data we can also extract the value of the contact angle hysteresis ($\Delta\theta$), i.e. the difference between the advancing and the receding contact angle. Again, for aligned fibers both the advancing and the receding angles are simultaneously measured in the direction parallel and perpendicular to the fiber length.

Finally, the maximum value of the volume ($V_{max}$) supported by the sample is measured by placing distilled-water drops of increasing volume on the horizontal surface and then tilting the surface vertically with a computer-controlled motor at a rate of ~1°/s.[41] Drops are deposited with a syringe pump (World Precision Instrument, Inc.) and the estimated uncertainty in $V$ is about 3%. Drops having $V$ larger than $V_{max}$ are found to fall down during the tilt, whereas drops with $V \leq V_{max}$ remain attached to the surface even though it is tilted fully upside down. For the aligned fibers, this procedure is applied twice, namely with the fibers either parallel or perpendicular to the gravity force. In spite of its conceptual simplicity, such a measurement is not common in published studies on sticky surfaces, which are mostly limited to determining contact and sliding angles.

## 3. Results

We have systematically varied the fiber diameter to explore the wetting behavior and, more importantly, the water adhesion to our electrospun coatings and non-wovens. Exemplary fibers from samples obtained with and without TBAI are shown in Fig. 1a,b (*1*-samples) and 1c,d (*3*-samples), respectively. All fibers exhibit a smooth and uniform surface. In Fig. 1, micrographs of randomly oriented fibers are reported in the top panels (Fig. 1a,c), whereas fibers in the bottom images form arrays prevalently aligning along their longitudinal axis (left-right axis in Fig. 1b,d).

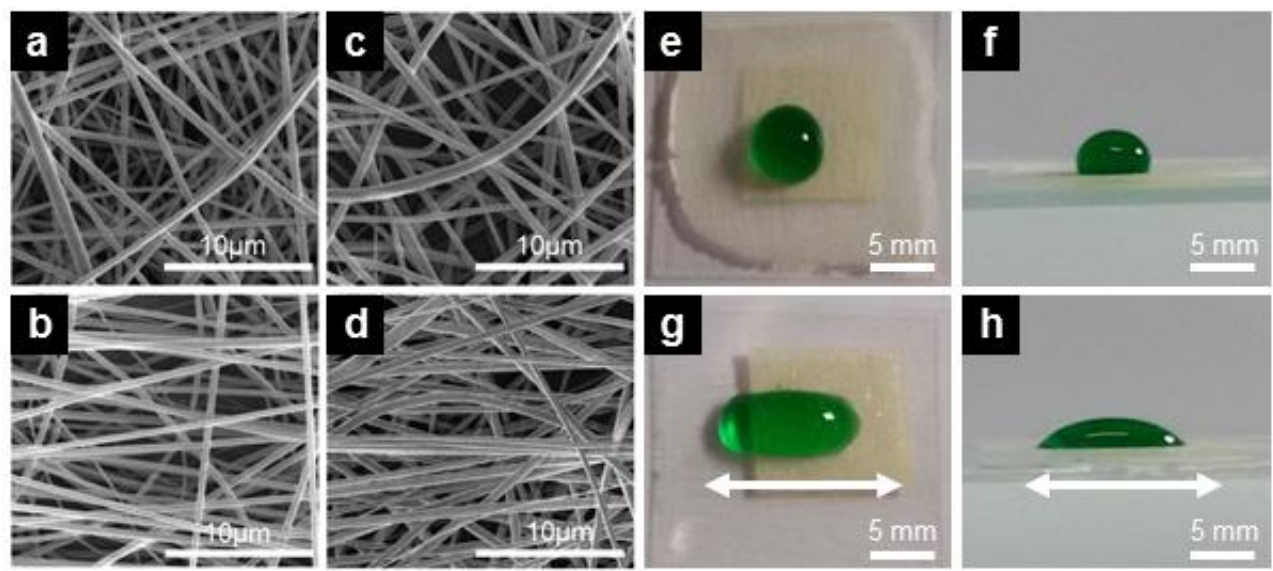

**Figure 1**. (a-d) SEM micrographs of fibers with different average diameter and orientation. (a) *1-R*, ϕ=(0.7±0.1) μm, (b) *1-A*, ϕ=(0.7±0.1) μm, (c) *3-R*, ϕ=(1.8±0.8) μm, (d) *3-A*, ϕ=(1.7±0.9) μm. (e-h) Colored water drops deposited on randomly oriented (e, f) and on aligned (g, h) fibers along the direction shown by the arrows in (g,h).

*Wetting behaviour*. The expectably different wetting behavior of random and aligned fibers is shown in Figure 1e-h, displaying a drop of green-colored water which assumes a circular contour with a contact angle significantly larger than 90° on a random (*R*) sample, and a markedly elongated shape along the fiber direction on an aligned (*A*) sample, with the contact angle depending on the orientation of the view.

| Sample ID | Fiber diameter (μm) | Static contact angle (degrees) | Advancing angle (degrees) | Receding angle (degrees) | Contact angle hysteresis (degrees) | Maximum volume (μL) |
|---|---|---|---|---|---|---|
| **Glass** | | 36±2 | 50±2 | 15±3 | 35±4 | 6±1 |
| **PMMA** | | 76±2 | 78±2 | 33±5 | 45±5 | 12±2 |
| **1-$R_g$** | 0.7±0.1 | 121±5 | 136±6 | 20±3 | 116±7 | 44±2 |
| **2-$R_g$** | 1.3±0.8 | 118±7 | 125±7 | 16±2 | 109±7 | 50±2 |
| **3-$R_g$** | 1.8±0.8 | 127±8 | 138±4 | 24±6 | 115±7 | 48±6 |
| **4-$R_g$** | 5.2±0.1 | 129±5 | 137±5 | 15±4 | 122±6 | 53±8 |
| **$R_{fs}$** | 1.3±0.8 | 141±4 | 146±7 | 28±9 | 118±11 | 40±2 |

**Table 1.** Wettability characterization of the randomly oriented fiber mats. The g and fs subscripts indicate mats deposited on glass or free standing, respectively. Values measured on pristine bare glass and PMMA are also shown for comparison. See text for more details.

The wettability characterization of $R_g$-samples formed by fiber mats deposited on glass is summarized in Table 1. For sake of comparison, and in order to evidence an eventual role played by the substrate underneath in determining the overall wettability properties, we have also characterized free standing mats with similar morphologies ($R_{fs}$-samples). At least two (typically four) specimens are analyzed for each batch. Fig. 2 displays the resulting apparent contact angle ($\theta$) and contact angle hysteresis ($\Delta\theta$) behavior, which are not found to vary significantly upon varying the fiber size from 0.7 to 5.2 μm. Overall, $\theta$ is comprised between 120° and 136°, increasing by more than 40° with respect to the value measured on a flat PMMA surface. These results are in agreement with previous results on poly(vinyl butyral) (PVB)[28] and on fluorinated polyimide fibers.[38] Compared with a smooth PVB surface prepared by spin coating, whose intrinsic contact angle is 60°, a mat of PVB fibers of ~0.6 μm in size exhibits a higher $\theta \sim 132°$. This angle increases monotonically to 143° as the fiber diameter is decreased to ~0.1 μm. Similarly, randomly deposited fluorinated polyimide fibers with diameters ranged from ~0.1 μm to ~0.5 μm are found to increase the water contact angle to 143° from an intrinsic value of about 100°.[38] However, their adhesive behavior is the opposite: PMMA fiber mats are sticky, PVB and fluorinated polyimide fiber mats are highly water repellent, because PMMA is hydrophilic, while the two other materials are hydrophobic.

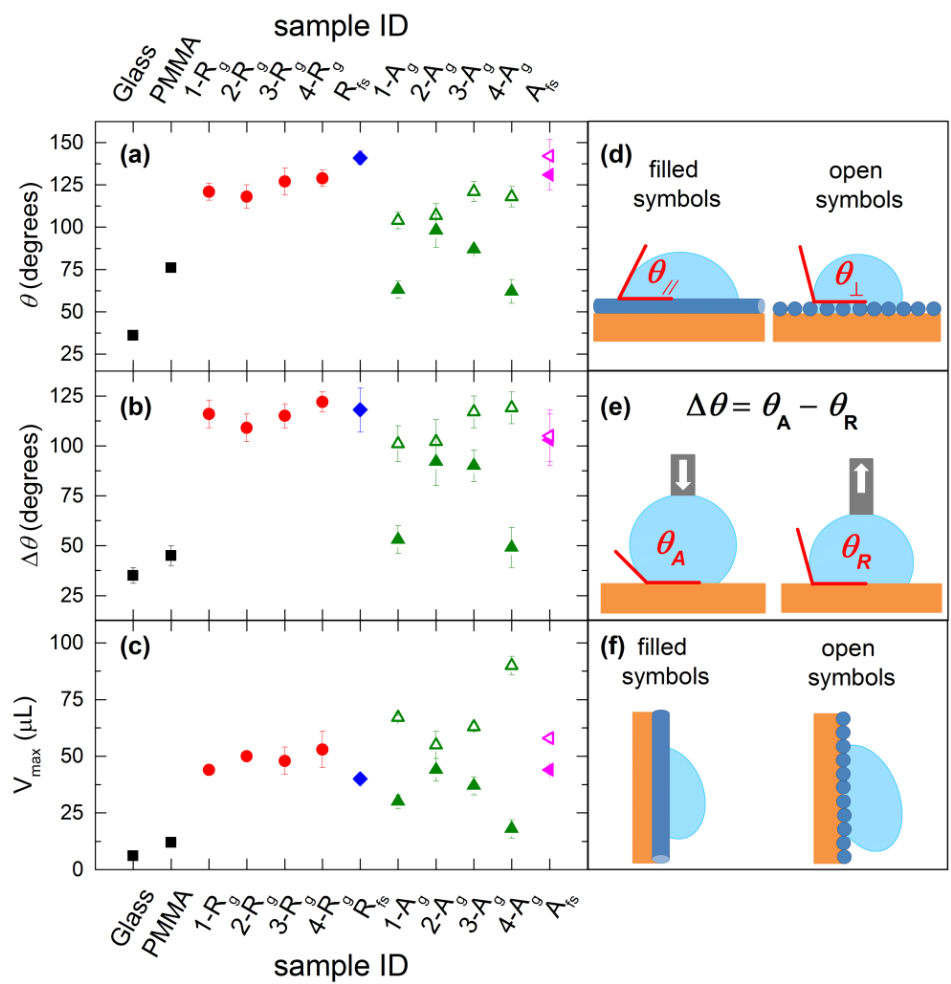

**Figure 2.** (a) Contact angle, (b, e) contact angle hysteresis and (c) maximum supported volume measured on flat substrates, randomly oriented fiber mats (R) and aligned fiber samples (A). For A-samples, filled (open) symbols refer to the direction parallel (perpendicular) to the oriented fibers, as defined in the cartoons (d, f). For more details see Table 1.

The wetting behavior observed with the PMMA fiber mats cannot be explained by the Wenzel model, which would predict a decreased $\theta$ following texturing of the hydrophilic PMMA, in contrast to what we observe. Similarly, the drop cannot be in the Cassie regime, with air fully entrapped in the surface textures and the drop sitting on the composite solid-air interface, because this state is characterized by high contact angles but low hysteresis (sliding angles).[1] Hence, our measurements suggests the occurrence of a so-called Cassie *impregnating* state, intermediate between the two previous regimes, with water filling the large voids of the texture but not the small ones. This is supported by AFM topographic maps of randomly deposited PMMA fibers (Figure 3), highlighting a mat surface which consists of a complex 3D architecture (Fig. 3a) where a tangle of fibers generates a multilayered structure with root mean square roughness of 0.8 µm. The network of fibers forming the outward layer, in most direct contact with the liquid, covers about 50% of the total surface area. The corresponding height profiles appear jagged with alternating peaks and valleys, with voids of various extension and peak-to-peak average distances roughly ranging between a few tens of nm and about 5 µm (Fig. 3b), depending on the fiber size and specific arrangement/position within the mat. Such morphology exhibiting multiscale and multilayer texturing is highly suitable to promote hybrid impregnating states.[26, 42]

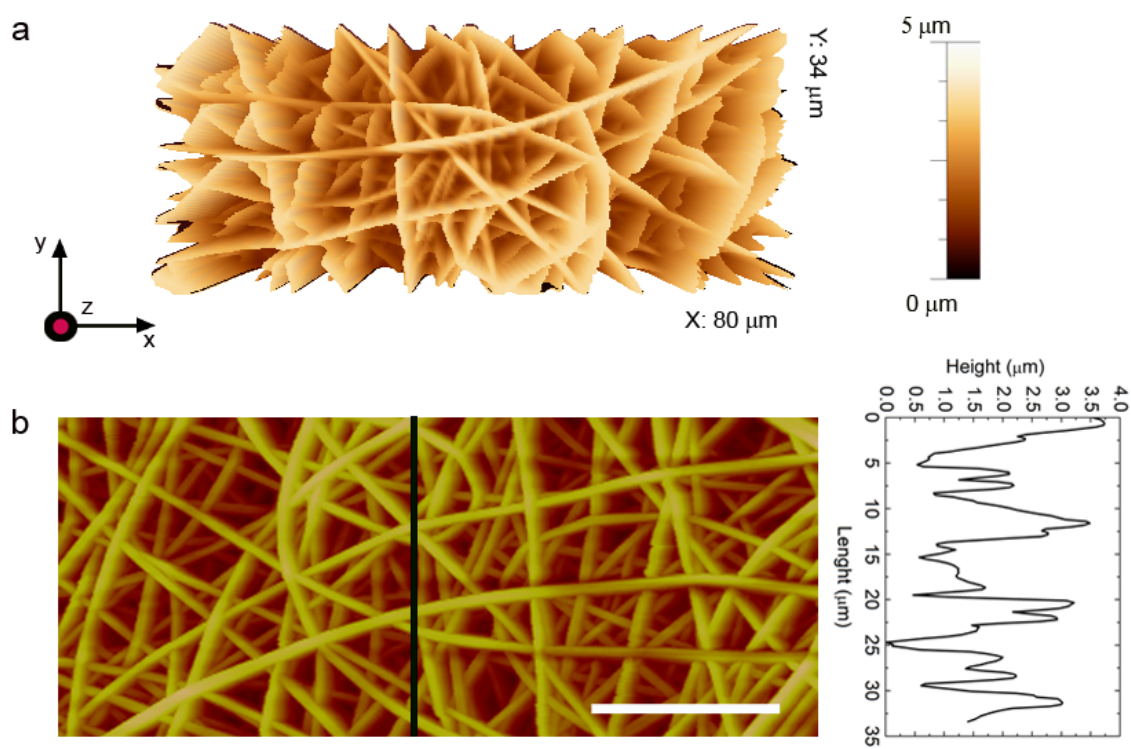

**Figure 3.** AFM topographic micrograph of a fiber mat (*2-R*$_g$). (a) 3D view. (b) 2D view. Scale bar: 20 µm. Vertical scale: 5 µm. The height profile is taken along the line highlighted in the planar view.

*Highly sticking behaviour*. Fiber surfaces present a similar apparent contact angle regardless of the fact that they are deposited on glass or free standing. Also, deposited mats present a very large contact angle hysteresis $\Delta\theta$ which makes them very sticky. This is confirmed by the maximum volume $V_{max}$ of the drops sustained by the surface tilted in a vertical position as shown in Figure 2c. The mats on glass and the free-standing carpets of nanofibers present $V_{max}$ ~45 μL or higher and $V_{max}$ ~40 μL, respectively, which corresponds to a spherical drop having a diameter of about 4.5 mm, almost twice the characteristic volumes obtained with hairy surfaces prepared in different ways.[8, 16, 17, 19, 43-46] The sticky behavior of these superhydrophobic surfaces is well-explained in terms of the Cassie impregnating state. The slightly lower $V_{max}$ value for free-standing samples might be related to minor topological differences in the electrospun fibers, which is reliable given the observed scatter in the data corresponding to samples prepared in the same nominal conditions. More importantly, data for free-standing mats indicate that the presence of the hydrophilic glass substrate plays a minor role in the wetting and adhesive forces of deposited coatings, at least for the dense mats realized in this study.

The images in Figure 4 provide further insight on the impregnation process. Figure 4a shows the profile of a sessile drop deposited on a randomly oriented fiber mat placed on glass. Figures 4b and 4c display the same drop after tilting the sample in the vertical position and upside down, respectively. At the beginning of plane tilting, the contact line is not completely pinned, and the front contact line moves downward due to the action of gravity and then gets pinned again, while the rear contact points are always pinned in the same position due to the very low receding contact angle. Figure 4d is especially interesting, presenting the drop contour after the sample is returned to the original horizontal position. It is evident an expansion of the contact area, which can increase up to almost 50%, and a corresponding decrease of the contact angle by about 20°-30°, which suggests that the initial profile was in a metastable regime and that the tilting of the sample favors a better impregnation of the fiber mat. We point out that, during plane tilting, the front angle increases and overcomes the advancing angle, while the rear angle decreases but it always remain higher than the receding angle. Consequently, only the front advances and the contact area becomes larger, implying a higher impregnation and adhesion with the surface. Further surface tilting does not alter this profile. Such an evolution is also observed with free-standing mats of nanofibers.

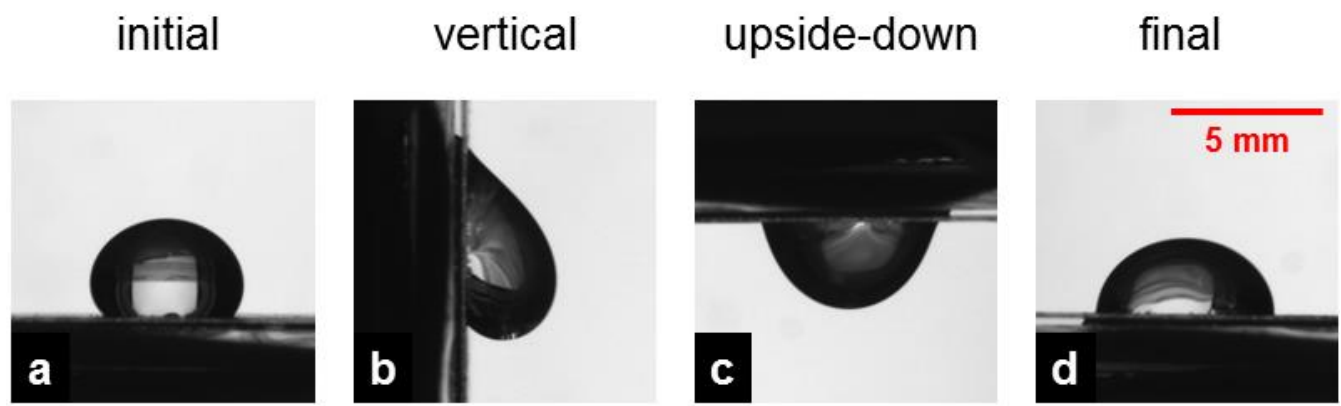

**Figure 4.** Typical drop profiles taken on randomly oriented fibers. (a) Sessile drop, (b) maximum volume held by the surface kept in the vertical position and (c) upside down, (d) the new drop contour after tilting the surface back to the horizontal position, exhibiting a lower apparent contact angle and a contact area larger than the initial one by about 40%. The scale is the same in all figures.

We have also studied mats with aligned fibers, whose wetting properties are summarized in Table 2 and in Figure 2. As mentioned above, the shape of water drops deposited on these surfaces is elongated in the direction of the fibers.[28, 47] While tending to expanding in every direction, a deposited drop finds randomly distributed energy barriers in its motion perpendicularly to the fiber longitudinal axis, while it can move along the parallel direction. For such drops the apparent static contact angles in the direction perpendicular or parallel to the fibers, $\theta_\perp$ and $\theta_{//}$, are therefore different (Fig. 2d), with $\theta_\perp > \theta_{//}$ as promoted by the relevant pinning phenomena. Defining $\delta\theta = \theta_\perp - \theta_{||}$ as the degree of wetting anisotropy, aligned fibers on glass present $\delta\theta \sim 40°$. The anisotropy is much less pronounced on the free-standing fiber mats, possibly suggesting that the fibers alignment is not so good as on the mats deposited on glass.

| Sample ID | Fiber diameter (µm) | Static contact angle (degrees) | | Advancing angle (degrees) | | Receding angle (degrees) | | Contact angle hysteresis (degrees) | | Maximum volume (µL) | |
|---|---|---|---|---|---|---|---|---|---|---|---|
| | | // | ⊥ | // | ⊥ | // | ⊥ | // | ⊥ | // | ⊥ |
| **1-$A_g$** | 0.7±0.1 | 63±5 | 104±5 | 77±6 | 119±6 | 25±4 | 19±7 | 53±7 | 101±9 | 30±3 | 67±1 |
| **2-$A_g$** | 1.2±0.3 | 98±10 | 107±7 | 107±11 | 119±11 | 15±4 | 17±2 | 92±12 | 102±11 | 44±5 | 55±6 |
| **3-$A_g$** | 1.7±0.9 | 87±2 | 121±6 | 112±8 | 132±7 | 23±2 | 15±4 | 90±8 | 117±8 | 37±4 | 63±2 |
| **4-$A_g$** | 4.6±1.2 | 62±7 | 118±6 | 71±9 | 136±6 | 22±4 | 17±5 | 49±10 | 119±8 | 18±4 | 90±4 |
| **$A_{fs}$** | 1.2±0.3 | 131±9 | 142±10 | 134±10 | 143±11 | 31±8 | 38±7 | 103±13 | 105±13 | 44±2 | >58±2 |

**Table 2.** Wettability characterization of aligned fibers. Labels follow the scheme adopted for samples featuring fibers randomly oriented (see Table 1), where *R* is replaced by *A*, standing for aligned fibers. See text for more details.

The anisotropy introduced by the aligned fibers also greatly affects $V_{max}$. When the fibers are aligned horizontally (i.e., perpendicular to the direction of the acceleration of gravity), they are able to sustain larger drops than when they are aligned vertically (see Fig. 2f). In particular, on *4-$A_g$* samples with horizontally aligned fibers we find a $V_{max}$ value as high as 90 µL, hardly reported before.

Sticky surfaces have been used to transfer water drops from a superhydrophobic surface to a hydrophilic one without any loss or contamination.[16, 43, 48] Figure 5 shows a water drop having a volume of 44 µL placed on a PDMS reproduction of a lotus leaf obtained by nanocasting,[13] with a contact angle ~150° and a sliding angle lower than 5°. Then, the drop is completely transferred to the fiber mat by simply touching the drop with the sample *3-$R_g$*. The drop is then released from the fiber mat to a glass slide. With a precision balance, we also quantify the transfer of water drops between different combinations of surfaces, with volume ranging between a few microliters and $V_{max}$. From glass coated with fiber mats to bare hydrophilic glass the transferred volume is more than 90% (practically the same) of the initial one.

The sticky behavior of coatings and surfaces based on electrospun polymer nanofibers is to be based on relevant Van der Waals forces, whose effect is enhanced by the high density of deposited non-wovens as well as by the large surface-to-volume ratio of the polymer filaments. We point out that the multilayered geometry of electrospun fibers is significantly different from previously reported, gecko-inspired pillar architectures.[16,43] Furthermore, the potentiality of electrospinning methods in view of realizing large-area samples is much higher than in other fabrication approaches, which opens interesting perspectives for various industrial applications. Fields benefiting from the capability of delivering fluids with reduced or no volumetric loss include analytical chemistry, microfluidics, catalytic and filtration technologies.

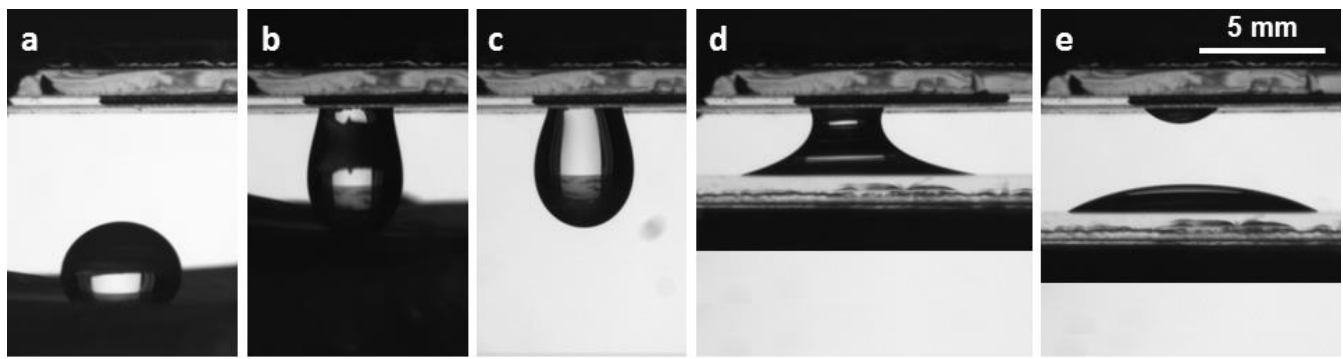

**Figure 5.** Transfer of a 44 µL water drop from a superhydrophobic water repellent sample to a glass hydrophilic surface through a glass coverslip coated with a fiber mat. (a) drop on the PDMS copy of a lotus leaf; (b) drop transfer from the lower superhydrophobic sample to the upper glass surface covered by a fiber mat; (c) drop pending from the fiber mat; (d) capillary bridge formed by the water drop between the upper fiber mat and the lower untextured glass slide; (e) drop deposited on the homogeneous glass slide. The scale is the same in all figures.

## Conclusions

We have studied the wetting and the unique adhesive properties of electrospun PMMA fibers. The fiber diameter has been intentionally varied in order to disentangle eventual fiber size-effects affecting the overall surface wettability behavior. On randomly oriented fibers, the apparent contact angle of deposited water drops is ~130°, regardless of the fibers size. More interestingly, very large adhesion forces capable to hold water drops as large as 60 µL, more than twice the characteristic values obtained with hairy surfaces, are achieved. Aligned fibers present anisotropic wetting behavior, and a maximum volume of water drops retained in the direction perpendicular to the fibers up to 90 µL. Measurements carried out on free-standing fiber mats indicate that the presence of the glass substrate plays a marginal role in determining the above mentioned features. This work suggests electrospun polymers as very promising tool to tailor surface wetting behavior, through up-scalable production of nanofibers which might exhibit modulated interactions with liquids.

## Acknowledgements

The research leading to these results has received funding from the European Research Council under the European Union's Seventh Framework Programme (FP/2007-2013)/ERC Grant Agreements N. 297004 ("Droemu") and n. 306357 ("NANO-JETS"). Grant PRAT 2011 "MINET" from the University of Padova and the Apulia Network of Public Research Laboratories Wafitech (9) are also gratefully acknowledged.

## References

1. P. G. De Gennes, F. Brochard-Wyart and D. Quéré, *Capillarity and Wetting Phenomena: Drops, Bubbles, Pearls, Waves*, Springer, New York, 2004.
2. E. Y. Bormashenko, *Wetting of real surfaces*, Walter de Gruyter, 2013.
3. B. Bhushan and Y. C. Jung, *Progress in Materials Science*, 2011, **56**, 1-108.
4. L. Feng, S. H. Li, Y. S. Li, H. J. Li, L. J. Zhang, J. Zhai, Y. L. Song, B. Q. Liu, L. Jiang and D. B. Zhu, *Advanced Materials*, 2002, **14**, 1857-1860.
5. K. S. Liu, X. Yao and L. Jiang, *Chemical Society Reviews*, 2010, **39**, 3240-3255.
6. B. Bhushan, *Langmuir*, 2012, **28**, 1698-1714.
7. W. Barthlott and C. Neinhuis, *Planta*, 1997, **202**, 1-8.
8. L. Feng, Y. Zhang, J. Xi, Y. Zhu, N. Wang, F. Xia and L. Jiang, *Langmuir*, 2008, **24**, 4114-4119.
9. X. M. Li, D. Reinhoudt and M. Crego-Calama, *Chemical Society Reviews*, 2007, **36**, 1350-1368.
10. P. Roach, N. J. Shirtcliffe and M. I. Newton, *Soft Matter*, 2008, **4**, 224-240.
11. T. Onda, S. Shibuichi, N. Satoh and K. Tsujii, *Langmuir*, 1996, **12**, 2125-2127.
12. J. Bico, C. Marzolin and D. Quere, *Europhysics Letters*, 1999, **47**, 220-226.

13. M. H. Sun, C. X. Luo, L. P. Xu, H. Ji, O. Y. Qi, D. P. Yu and Y. Chen, *Langmuir*, 2005, **21**, 8978-8981.
14. A. Pozzato, S. Dal Zilio, G. Fois, D. Vendramin, G. Mistura, M. Belotti, Y. Chen and M. Natali, *Microelectronic Engineering*, 2006, **83**, 884-888.
15. Y. H. Cui, A. T. Paxson, K. M. Smyth and K. K. Varanasi, *Colloids and Surfaces a-Physicochemical and Engineering Aspects*, 2012, **394**, 8-13.
16. M. H. Jin, X. J. Feng, L. Feng, T. L. Sun, J. Zhai, T. J. Li and L. Jiang, *Advanced Materials*, 2005, **17**, 1977-1981.
17. X. Wang and R. A. Weiss, *Langmuir*, 2012, **28**, 3298-3305.
18. J. B. K. Law, A. M. H. Ng, A. Y. He and H. Y. Low, *Langmuir*, 2014, **30**, 325-331.
19. S. Varagnolo, N. Basu, D. Ferraro, T. Toth, M. Pierno, G. Mistura, G. Fois, B. Tripathi, O. Brazil and G. L. W. Cross, *Microelectronic Engineering*, 2016, **161**, 74-81.
20. D. H. Reneker and I. Chun, *Nanotechnology*, 1996, **7**, 216-223.
21. D. Pisignano, *Polymer nanofibers*, Royal Society of Chemistry, Cambridge, 2013.
22. S. Agarwal, A. Greiner and J. H. Wendorff, *Progress in Polymer Science*, 2013, **38**, 963-991.
23. D. H. Reneker and A. L. Yarin, *Polymer*, 2008, **49**, 2387-2425.
24. L. Jiang, Y. Zhao and J. Zhai, *Angewandte Chemie-International Edition*, 2004, **43**, 4338-4341.
25. M. L. Ma, R. M. Hill, J. L. Lowery, S. V. Fridrikh and G. C. Rutledge, *Langmuir*, 2005, **21**, 5549-5554.
26. A. Tuteja, W. Choi, M. Ma, J. M. Mabry, S. A. Mazzella, G. C. Rutledge, G. H. McKinley and R. E. Cohen, *Science*, 2007, **318**, 1618-1622.
27. J.-M. Lim, G.-R. Yi, J. H. Moon, C.-J. Heo and S.-M. Yang, *Langmuir*, 2007, **23**, 7981-7989.
28. H. Wu, R. Zhang, Y. Sun, D. Lin, Z. Sun, W. Pan and P. Downs, *Soft Matter*, 2008, **4**, 2429-2433.
29. D. Han and A. J. Steckl, *Langmuir*, 2009, **25**, 9454-9462.
30. P. Muthiah, S.-H. Hsu and W. Sigmund, *Langmuir*, 2010, **26**, 12483-12487.

31. B. Grignard, A. Vaillant, J. de Coninck, M. Piens, A. M. Jonas, C. Detrembleur and C. Jerome, *Langmuir*, 2011, **27**, 335-342.

32. J. P. Zhang and S. Seeger, *Advanced Functional Materials*, 2011, **21**, 4699-4704.

33. J. P. Zhang and S. Seeger, *Angewandte Chemie-International Edition*, 2011, **50**, 6652-6656.

34. S. J. Hardman, N. Muhamad-Sarih, H. J. Riggs, R. L. Thompson, J. Rigby, W. N. A. Bergius and L. R. Hutchings, *Macromolecules*, 2011, **44**, 6461-6470.

35. X. P. Tian, L. M. Yi, X. M. Meng, K. Xu, T. T. Jiang and D. Z. Lai, *Applied Surface Science*, 2014, **307**, 566-575.

36. T. Pisuchpen, N. Chaim-ngoen, N. Intasanta, P. Supaphol and V. P. Hoven, *Langmuir*, 2011, **27**, 3654-3661.

37. B. S. Lalia, S. Anand, K. K. Varanasi and R. Hashaikeh, *Langmuir*, 2013, **29**, 13081-13088.

38. G. M. Gong, J. T. Wu, Y. Zhao, J. G. Liu, X. Jin and L. Jiang, *Soft Matter*, 2014, **10**, 549-552.

39. D. Ferraro, C. Semprebon, T. Toth, E. Locatelli, M. Pierno, G. Mistura and M. Brinkmann, *Langmuir*, 2012, **28**, 13919-13923.

40. T. Toth, D. Ferraro, E. Chiarello, M. Pierno, G. Mistura, G. Bissacco and C. Semprebon, *Langmuir*, 2011, **27**, 4742-4748.

41. S. Varagnolo, D. Ferraro, P. Fantinel, M. Pierno, G. Mistura, G. Amati, L. Biferale and M. Sbragaglia, *Physical Review Letters*, 2013, **111**, 066101.

42. M. Moffa, A. Polini, A. G. Sciancalepore, L. Persano, E. Mele, L. G. Passione, G. Potente and D. Pisignano, *Soft Matter*, 2013, **9**, 5529-5539.

43. W. K. Cho and I. S. Choi, *Advanced Functional Materials*, 2008, **18**, 1089-1096.

44. E. Bormashenko, T. Stein, R. Pogreb and D. Aurbach, *Journal of Physical Chemistry C*, 2009, **113**, 5568-5572.

45. J. Peng, P. Yu, S. Zeng, X. Liu, J. Chen and W. Xu, *Journal of Physical Chemistry C*, 2010, **114**, 5926-5931.

46. X. Liu, Q. Ye, B. Yu, Y. Liang, W. Liu and F. Zhou, *Langmuir*, 2010, **26**, 12377-12382.

47. G. Morello, R. Manco, M. Moffa, L. Persano, A. Camposeo and D. Pisignano, *ACS Applied Materials & Interfaces*, 2015, **7**, 21907-21912.

48. C. F. Wang and T. W. Hsueh, *Journal of Physical Chemistry C*, 2014, **118**, 12399-12404.